Chemically driven isothermal closed space vapor transport of $MoO_2$: thin films, flakes and in-situ tellurization

O. de Melo,[a,b] L. García-Pelayo,[a] Y. González,[a,b] O. Concepción,[a,c] M. Manso-Silván,[a] R. López-Nebreda,[a] J. L. Pau,[a] J. C. González,[d] A. Climent-Font,[a,e] V. Torres-Costa[a,e]

[a]Departamento de Física Aplicada e Instituto de Ciencia de Materiales Nicolás Cabrera, Facultad de Ciencias, Universidad Autónoma de Madrid, Cantoblanco 28049, España

[b]Physics Faculty, University of Havana, 10400 La Habana, Cuba

[c] Nanoscience and Nanotechnology PhD Program, Cinvestav-IPN, 07360 Mexico City, Mexico

[d]Departamento de Física, Universidad Federal de Minas Gerais, Belo Horizonte, MG 30123-970, Brazil

[e]Centro de Microanálisis de Materiales. Universidad Autónoma de Madrid, 28049, Madrid, Spain

Abstract.

A novel procedure, based in a closed space vapor transport (CSVT) configuration, has been devised to grow films or flakes of pure $MoO_2$ in a reductive atmosphere, at relatively low temperature and using $MoO_3$ as the source. In contrast with conventional CSVT technique, in the proposed method a temperature gradient is not required for the growth to take place, which occurs through an intermediate volatile transport species that is produced in the complex reduction reaction of $MoO_3$. An added value of this simple method is the possibility of transforming the $MoO_2$ into $MoTe_2$, one of the most interesting members of the transition metal dichalcogenide family. This is achieved in a sequential process that includes the growth of Mo oxide and its (in-situ) tellurization in two consecutive steps.

1. Introduction

Molybdenum is a relatively abundant transition metal, which forms different oxide modifications with formula $MoO_x$ and *x* ranging from 2 to 3. The reported compounds $MoO_3$, $Mo_{18}O_{52}$, $Mo_{17}O_{47}$, $Mo_9O_{26}$, $Mo_8O_{23}$, $Mo_5O_{14}$, $Mo_4O_{11}$ and $MoO_2$ belong to this family. The orthorhombic structure is the most stable form of $MoO_3$, which has n-type conductivity and a band gap of around 3.0 eV.[1] Meanwhile, $MoO_2$ has the monoclinic structure as its more stable phase,[2] but its electrical properties are somewhat uncertain. Although theoretical calculations predict a metallic behavior as a consequence of Fermi level crossing by Mo 4d bands,[3] an optical band gap has been found in absorption spectra.[4] An interesting account for molybdenum oxides properties and applications can be found in ref. 5.

Both oxides are very promising materials. $MoO_3$ has been used in catalysis,[6,7] chemical sensing,[8,9] energy storage,[10] and electrochromism[11] among other applications. $MoO_2$ has been also identified as catalyst[12,13] and, due to its strong near infrared absorption, it has been successfully tested in photo-thermal therapies.[14] $MoO_2$ has been used alternatively as anode material in lithium ion batteries[15] and as electrode for supercapacitors.[16] Applications of $MoO_2$ in energy conversion and storage have been summarized by Y. Zhao et al.[17] Prominently, both $MoO_3$ and $MoO_2$ are commonly used as precursors for the preparation of 2D Mo dichalcogenides.

Regarding $MoO_x$ processing, recent research efforts have been directed to its synthesis through different routes such as chemical vapor deposition,[18] $MoO_3$ reduction,[19] electrospinning,[20] hydrothermal process,[12] pulsed laser deposition,[21] thermal evaporation[22] and reactive sputtering.[23]

Closed space sublimation (CSS) (or the more general appellative, closed space vapor transport, CSVT) is a relatively simple and cost effective growth technique which has been used for the preparation of thin films of several materials in both epitaxial and polycrystalline form. The most widespread example of the use of this technique is the growth of CdTe films for CdTe based solar cells,[24] but other materials like $Bi_2Te_3$[25] or epitaxial GaAs[26] and GaP[27] have been also prepared with this method. Typically, source material and substrate are located in a semi-closed environment, the substrate above, a few mm away and at lower temperature than the source. In this configuration, the source-substrate temperature gradient is the driving force for the thin film growth: a forward reaction is favored in the source while the reverse reaction occurs at the growing surface. In the case of materials that sublimate congruently, such as CdTe, the forward and reverse reactions are simply sublimation and condensation, respectively. If this is not the case, a reactive vapor can be supplied in the growth compartment to help in the transport of the non-volatile source component. For example, for the growth of GaAs by CSVT, water vapor has been introduced to favor the forward reaction $GaAs\ (s) +\ H_2O\ (g) \rightarrow Ga_2O\ (g) + As_2\ (g)$ at the source and the reverse reaction at the growing surface.[26]

In this paper, we report the isothermal CSVT growth of pure $MoO_2$ in a reducing atmosphere using $MoO_3$ powder as the source. In contrast to previously reported CSVT processes, the growth mechanism does not involve the same reaction occurring in the forward/reverse direction at the source/substrate, since this would be impossible in isothermal conditions. Instead, the method presented here follows a novel chemically driven process based on the complex reduction of $MoO_3$: the first step occurring at the source and the final step at the substrate, a few millimeters away. Thus, the configuration used here promotes that the final $MoO_2$ product will appear distant from the initial $MoO_3$ reactive powder. This very simple method has the advantage of producing pure $MoO_2$ films (that can be converted to $MoO_3$ or other desired $MoO_x$ compound by simple annealing procedures) with well controlled thickness and homogeneity, contrary to other growth procedures in which a mix of different $MoO_x$ phases is often obtained.

The objective of the present work has been to monitor the growth of $MoO_2$ along a wide range of growth conditions in order to determine the CSVT growth dynamics. Moreover, a particularly useful application, the in-situ tellurization of $MoO_2$ films, has been proposed with this method. This combined chemically driven (CD) CSVT and tellurization process should be an efficient alternative to the typically used procedure that involves sputtering of Mo or $MoO_x$ and tellurization in two separated steps. In order to assess the viability of the CD-CSVT method presented here, the morphological, (micro)structural and compositional properties of $MoO_2$ and $MoTe_2$ samples grown by this technique have been studied.

2. Experimental methods.

A scheme of the graphite boat used for the growth of $MoO_2$ is shown in Fig. 1a). It consists of a lower part containing the $MoO_3$ powder source and an upper substrate holder that can slide over the source by means of a rod, which is fed through the lateral cap of the reactor. For in-situ growth/tellurization experiments, a graphite boat with an additional container for the Te source was used. The boat is located in the temperature plateau region of a quartz reactor heated by a resistive tubular furnace, which assures a homogenous temperature over the whole graphite crucible, as measured in previous experiments.[28,29] (The temperature dynamics measured at opposite sides of the crucible during the growth process is shown in the supplementary material. See Fig. S.1). This reactor is flowed with a reducing gas mixture composed of $H_2$:Ar in proportion 1:5 at a rate of 25 mL/min and at atmospheric pressure. $SiO_2$ (300 nm)/Si or Si (100) were used as substrates. The growth temperature was varied between 570 and 630 $^0$C, the growth time between 10 and 90 min and the source-substrate distance ($d_{s\text{-}s}$) between 2 and 9 mm. In order to calibrate different $d_{s\text{-}s}$ without changing the total amount of $MoO_3$ reactive powder, small graphite disks with appropriate compensating heights were placed at the bottom of the source compartment.

To determine the crystalline structure of the samples, grazing incidence (0.5°) x-ray diffraction (XRD) scans were performed using a Siemens D-5000 powder diffractometer with $CuK_{\alpha 1}$ radiation. High resolution scanning electron microscopy (HR-SEM) images were obtained by using a FE-SEM Hitachi S-4700 microscope. Micro-Raman spectra were taken through a Bruker-Senterra spectrometer using a 50x objective and a 10 mW 532 nm excitation laser with resolution between 3-5 cm$^{-1}$. Ion backscattering spectroscopy (IBS) was performed with a 3.07 MeV α-particle beam provided by the Cockcroft-Walton tandem accelerator in order to increase the sensitivity to oxygen. IBS spectra were simulated using the SIMNRA code[30] in order to determine the composition and thickness of the samples. Height profiles were obtained with a Veeco Dektak 150 Profilometer. X-ray photoelectron spectroscopy (XPS) was carried out in a SPECS PHOIBOS 150 9MCD. Spectra were obtained by irradiating with a non-monochromatic Al K$_\alpha$ source and normal take off angle acquisition with pass energy of 75 eV for survey spectra and 25 eV for core level and valence band (VB) spectra. Casa XPS software was used for data analysis using Gaussian/Lorentzian fractions of 30, Shirley baseline and core level energies referred to the C1s line at 285.0 eV. Area constrictions were imposed to ensure spin orbit splitting intensity ratios were fulfilled in the Mo3d core level peaks. Atomic force microscope (AFM) images were obtained in contact mode with a Natotec Electronica system and processed with the WSxM software.[31]

2. Results and discussion

a. Morphology and phase identification

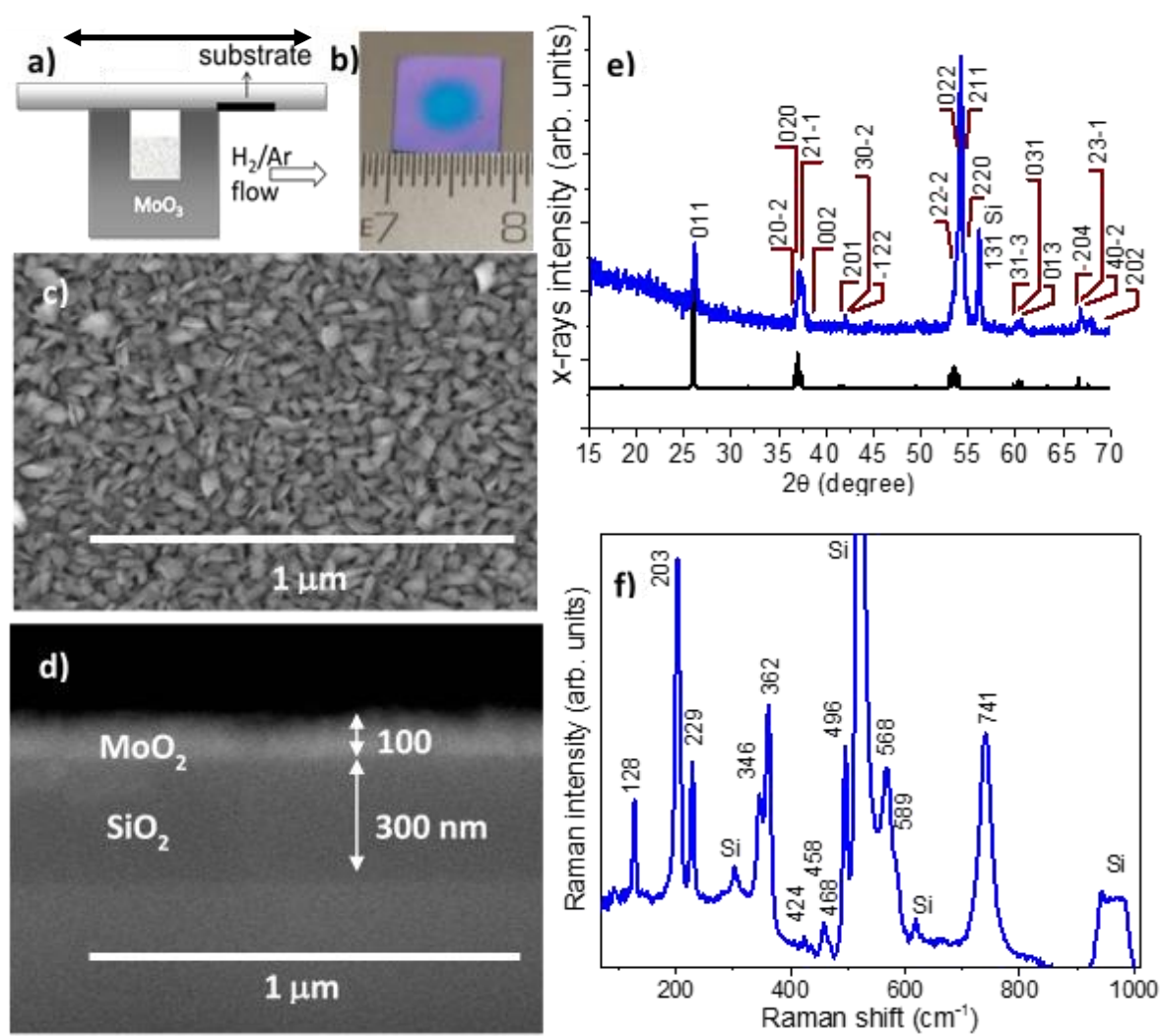

Fig. 1. a) Scheme of the graphite boat used for the isothermal closed space vapor transport growth of MoO$_2$; b) A typical MoO$_2$ film grown onto a SiO$_2$/Si substrate showing the optical contrast

between the film and the substrate. Scanning electron micrograph of c) the surface and d) the cross section of a MoO$_2$ film. e) Grazing incidence x-ray diffractogram. All peaks are indexed according to the MoO$_2$ monoclinic structure (except that originating in the Si substrate). In black, the reference pattern for the MoO$_2$ monoclinic structure (P121/c1 or group 14 system); f) Raman spectra for the same sample, all the peaks corresponding to MoO$_2$ except those coming from the Si substrate.

The naked eye aspect of a 15 nm thick MoO$_2$ sample is shown in Fig. 1b). This sample was grown at 570 °C for 10 min, with a $d_{s-s}$ of 4 mm onto a SiO$_2$/Si substrate, which allows a very good optical contrast between the MoO$_2$ film (the inner circle) and the 300 nm thick SiO$_2$ layer (external square). HR-SEM images of the surface and cross section of a 100 nm MoO$_2$ film are shown in c) and d), respectively. The top view image shows that the film is composed of flaky or lamellar confluent structures denoting a high surface roughness. The cross section image confirms this observation and provides a clear picture of the final thicknesses of the MoO$_2$ and SiO$_2$ in the structure.

The diffractogram of a typical MoO$_2$ film grown by CD-CSVT technique is shown in Fig. 1e). Except for the (131) peak of Si, commonly observed in grazing incidence measurements of (001) Si substrates, the rest of the peaks can be indexed according to the monoclinic structure (P121/c1 or group 14 system, PDF number 01-073-1249) which is the most stable structure of MoO$_2$ (a reference diffractogram is also shown in the figure). Additionally, the Raman spectrum of a typical sample is shown in Fig. 1f). The observed peaks are similar to that identified by other authors for MoO$_2$.[32,33] It can be noted that peaks in the range 650 - 1000 cm$^{-1}$, typical of Mo$_4$O$_{11}$ and MoO$_3$, are not observed in our spectra, confirming, aside the diffraction data, that the as-grown layers are pure MoO$_2$.

The IBS spectrum of a sample grown onto a SiO$_2$ (300 nm)/Si substrate is shown in Fig. 2 in which the surface positions of the different components are marked with arrows. Simulation of the spectra by SIMNRA allows to calculate the composition and thickness. Fitting of the spectra yields molar fractions of 0.34 and 0.66 for elemental Mo and O, which are very close values to the expected composition of MoO$_2$, and is in good agreement with XRD and Raman results. Calculated thickness by this procedure was 42 nm in this case. Data and procedure for determining the thickness from the IBS spectrum as well as the spectrum of a different sample are presented as supplementary material (see Fig. SII).

Although x-ray diffractograms and Raman spectra indicate unambiguously the presence of pure crystalline $MoO_2$, XPS measurements demonstrated that, at surface level, the atmospheric exposure of the synthesized layers induced an aging of the surface composition. The analysis of the aging process by XPS (illustrated in Fig. 3) showed that the transformation consisted of an oxidation of the surface of $MoO_2$ into $MoO_3$. To avoid modification of the aged films, no surface treatment was applied prior to analysis, which justified a considerable intensity of the C1s component in the survey spectra (below 25 at. % in all the spectra) due to atmospheric contamination. The estimation of atomic concentrations demonstrated that the aging effects are acute in as-grown $MoO_2$ films. While the general surface stoichiometry for samples measured one day after growth still preserves the bulk stoichiometry (Mo/O= 0.48), the surface composition drastically evolves in favor of a $MoO_3$ phase (Mo/O= 0.38) for samples measured one week later.

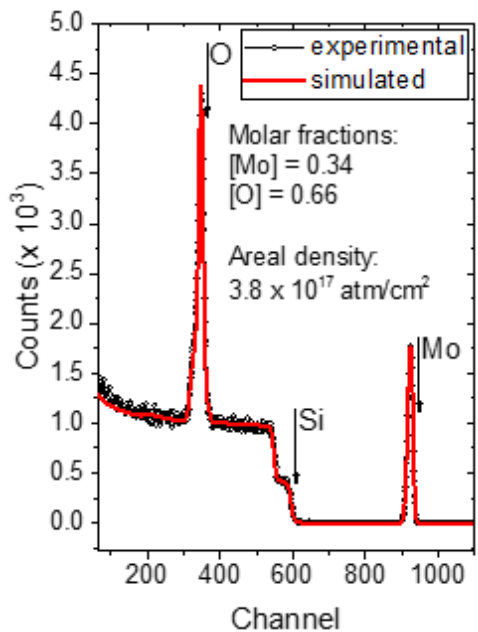

Fig. 2. Rutherford backscattering spectrum for a Mo oxide film grown onto a $Si/SiO_2$ substrate. From the simulation of the spectrum, the elemental composition Mo:O was found to be very near to the expected 1:2 corresponding to $MoO_2$.

Mo3d core level spectra were acquired for two different samples: one day after the growth process, and stored for one week prior to analysis. The spectra are shown in Figs. 3a) and b), respectively. A two phase deconvolution model was applied revealing the presence of MoO$_3$ and MoO$_2$, with increasing fraction of MoO$_3$ at longer atmospheric exposure. In fact, to fulfill the fitting, we used the first component observed at lower binding energy (BE) (230.4 eV) as the Mo3d$_{5/2}$ contribution due to MoO$_2$ phase. Since spin orbit splitting imposes constrictions in the area of contributions related by their orbital, the Mo3d$_{3/2}$ peak at 233.8 eV was accordingly constricted to 2/3 of the area (5% margin) of the Mo3d$_{5/2}$ contribution. This procedure allowed the estimation of the peak generated by the Mo3d$_{5/2}$ contribution of MoO$_3$, whose BE (233.3 eV) overlaps with the Mo3d$_{3/2}$ peak due to MoO$_2$. A similar restriction was imposed to the related areas of MoO$_3$ spin orbit peaks, with final Mo3d$_{3/2}$ contribution at 236.6 eV. These deconvolution analyses allowed establishing a ratio of surface phases, confirming that the MoO$_3$ phase progresses steadily with aging time (MoO$_3$/MoO$_2$ ratio from 1.7 to 2.7). The apparent contradiction among data stemming from estimation of the general surface stoichiometry (wide scan) and estimation of the phase ratio (core levels) suggests a relevant influence of the surface phase on the affinity to adsorb more or less oxidized carbon species.

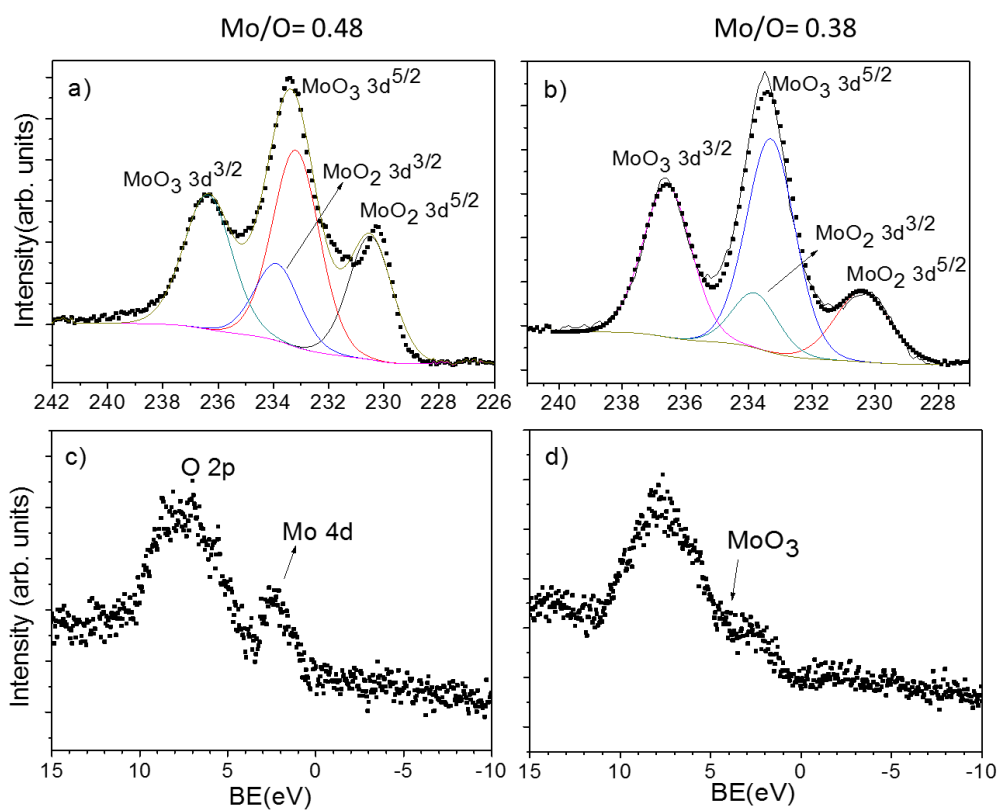

Fig. 3. a) Mo3d core level spectrum of a MoO$_2$ sample aged for 24 h. b) Mo3d core level spectrum of a MoO$_2$ sample aged for one week. c) BV spectrum of a MoO$_2$ sample aged for 24 h. d) BV spectrum of a MoO$_2$ sample aged for one week.

The increasing presence of the MoO$_3$ phase upon aging was observed to have a direct influence in the modification of the VB spectra of the MoO$_2$ samples. The VB for the sample measured after growth, Fig. 3 c), was observed to be in agreement with that reported for MoO$_2$ phase.[34] An emerging double peak (two shoulders) appears at BEs below 2 eV due to Mo4d orbital contribution. A second wider and more intense peak is observed at 8 eV as a contribution of the O2p orbital. The increasing aging has a direct effect in the low BE contributions with a neat decrease of the states observed at low binding energies and an increasing population through a new band arising at circa 3 eV due to the MoO$_3$ phase[35] as shown in Fig. 3d).

b.  MoO$_2$ isothermal growth mechanism

While in the vast majority of cases a temperature gradient acts as the driving force for CSVT growth, cases of growth in isothermal configuration have been also reported. For example, Hg$_x$Cd$_{1-x}$Te films have been grown with this configuration (with the name of Isothermal Vapor Phase Epitaxy, ISOVPE).[36,37] In this case, source and substrate are HgTe and CdTe, respectively, and Cd-Hg interdiffusion sustains, during growth, a Hg$_x$Cd$_{1-x}$Te surface. Then, the lower vapor pressure at the substrate than at the source provides the necessary chemical potential difference without the need of a temperature gradient. Other example is the Isothermal Closed Space Sublimation (ICSS)[28] technique in which alternated exposure to different elemental sources provides the necessary driving force to form the desired compound at the substrate. Some II-VI semiconductors and alloys have been grown with this method.[29] However, in the isothermal closed space vapor transport growth of molybdenum oxide presented here, the growth takes place without the help of any of the above mentioned mechanisms for ISOVPE or ICSS.

Although not in a closed space geometry, isothermal growth of MoO$_2$ from MoO$_3$ has been previously reported by heating SiO$_2$/Si substrates embedded in MoO$_3$ powder under a reducing atmosphere.[38] The authors of this growth method proposed a mechanism according to which H₂O, produced by the reduction of MoO₃ with H₂, reacts in turn with MoO₃ resulting in the volatile MoO₃(OH)₂ compound. Then, this intermediate species reacts with H₂ to give MoO₂ onto the embedded SiO₂ surface.

On the other hand, it has also been proposed[39] that during the reduction of MoO₃ to MoO₂, besides the pseudomorphic transformation, a chemical vapor transport mechanism is present in which a volatile intermediate specie (not specified in that report) promotes the nucleation of MoO₂ resulting in a completely new morphology. Moreover, it has been recognized[40] that one of the mechanisms of volatilization of MoO₃ in humid environments is the formation of MoO₂(OH)₂, a volatile specie whose vapor pressure exceeds that of the MoO₃ for high water vapor content. On the other hand, a study of the room temperature reduction of MoO₃ using the membrane approach for high pressure XPS, demonstrated that intercalated hydrogen forms H$_x$MoO₃ as an intermediate product, which slowly decomposes to form MoO₂ and water.[41]

All the above reports indicate that the reduction of MoO₃ is mediated by the formation of some hydrated molybdenum oxide compound (MoO₃(OH)₂, MoO₂(OH)₂ or H$_x$MoO₃) which is a volatile specie that can be further reduced to form MoO₂. In view of these previous observations, we explain the isothermal CSVT growth of MoO₂ by considering that the volatile hydrated specie produced during the reduction of the MoO₃ powder can diffuse toward the surface of the powder, where It can be then transported in the gas phase, react with H₂ and lead to the nucleation of MoO₂ onto the substrate.

According to these assumptions, two reactions occur in the MoO₃ source:

$$\text{MoO}_3(s) + \text{H}_2(g) \rightarrow \text{MoO}_2(s) + \text{H}_2\text{O}(g) \qquad (1)$$

$$\text{MoO}_3(s) + \text{H}_2\text{O}(g) \rightarrow \text{VS}(g) \qquad (2)$$

In which VS stands for the volatile specie, which plays the role of precursor for the MoO₂ growth. Eq. 1 is responsible for H₂O production, necessary for the completion of the second reaction (eq. 2). In this stage, the VS is produced.

In the substrate surface, a third reaction occurs which is responsible for MoO₂ growth:

$$\text{VS (g)} + \text{H}_2(\text{g}) \rightarrow \text{MoO}_2(\text{s}) + \text{H}_2\text{O(g)} \tag{3}$$

After several growth experiments at different conditions, some trends were observed. The presence of a reductive atmosphere containing $H_2$ was indispensable for the $MoO_2$ growth; this agrees with the above proposed mechanism in which $H_2$ consumption is expected both at the source and at the growing surface. The amount of $H_2$ flowing in the growth compartment through the small gap between the substrate surface and the graphite on which it is supported, was found to be enough for the growth process completion. $Mo_4O_{11}$ has been frequently observed as an intermediate compound in the oxidation process from $MoO_3$ to $MoO_2$. Probably, the reduction path occurring in our conditions, through a volatile specie, excludes the formation of that compound and our as-grown films are composed of pure $MoO_2$.

Experiments were performed to study the effect of $d_{s\text{-}s}$ on the growth rate of $MoO_2$. The thickness (measured with the profilometer) dependence on $d_{s\text{-}s}$ is shown in Fig. 4 for samples grown at 600 °C for 1 h in a $d_{s\text{-}s}$ range from 2 to 7 mm. For larger distances, the thickness resulted very small to be measured with the profilometer and, in general, the films did not cover the substrate surface completely. The characteristic aspect of the films in the different $d_{s\text{-}s}$ ranges and the corresponding height profiles measured along the diameter are also displayed in the figure. The thickness dependence on $d_{s\text{-}s}$ indicates the prevalence of the source to substrate diffusive vapor transport as the limiting growth factor for $d_{s\text{-}s}$ of 3 mm or larger.

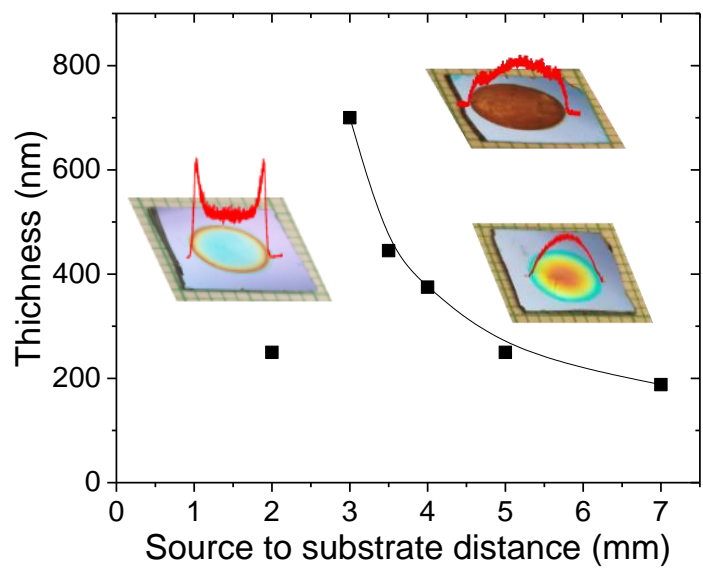

Fig. 4. Dependence of the film thickness with the source to substrate distance for samples grown for 1 h at 600 °C. Images and height profiles of different regimes are shown in the insets. (the line is only for guiding the observation of the dependence)

The VS was found to be very reactive and deposited also on the walls of the graphite compartment. This contributes to a decrease of the growth rate towards the border of the films (closer to the graphite walls) which was particularly important for larger $d_{s-s}$. This resulted in a radial thickness profile and produced convex $MoO_2$ films (see insets in figure 4). On the contrary, for $d_{s-s}$ smaller than 3 mm, an important increase of the growth rate was observed just in the border of the films, close to the graphite walls. This suggests that, in this regime, for small $d_{s-s}$ values, the growth rate is no longer limited by the vapor transport from the source, but by the amount of $H_2$ entering into the compartment through the gap between the substrate and the graphite boat. Hence $MoO_2$ grows predominantly near the $H_2$ entrance. This excessive deposition around the substrate border seems to hinder the net inwards $H_2$ flow, leading to a decrease of the film thickness at the substrate center despite the small $d_{s-s}$. In this regime, a concave radial thickness profile is obtained. Intermediate $d_{s-s}$ values of around 3-4 mm provided films relatively homogeneous in thickness. An additional experiment was performed to confirm the above described issues.

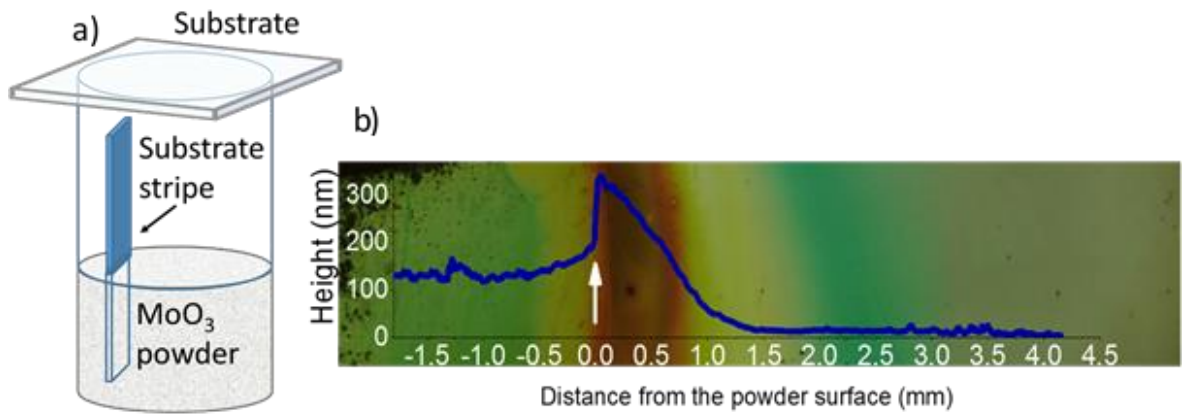

Fig. 5. a) Scheme of the growth system used to compare the growth process occurring inside and outside the MoO$_3$ powder; b) optical microscope view of the stripe with the corresponding superimposed film thickness profile.

Since MoO$_2$ is known to grow on the surface of a substrate embedded in the MoO$_3$ powder,[36] it is interesting to compare the deposition process occurring inside and outside the MoO$_3$ powder. To make such a comparison in a unique growth run, we buried the lower part of a SiO$_2$/Si substrate stripe in the MoO$_3$ powder while letting the uncovered upper part to emerge over the powder surface as indicated in Fig. 5a). The top of the compartment was closed with a substrate arranged as for the normal growth experiments. Growth temperature and time were 570 °C and 1h, respectively.

The results of this experiment can be resumed as follows. MoO$_2$ did not grow on the top substrate covering the container, nor in upper extreme region of the stripe. This indicates that, along a few millimeters, growth on the stripe consumed all the precursor generated in the powder. Fig. 5b) shows an optical microscope view of the stripe after MoO$_2$ growth with the corresponding film thickness profile superimposed. The boundary between the embedded (left) and the uncovered (right) regions of the stripe is marked with a white arrow. As can be observed, MoO$_2$ grew in both regions but, interestingly, an abrupt thickness increase is observed in the boundary: the growth rate was larger in the upper side of the border where the substrate was outside the powder. This observation can be explained by the different diffusion coefficients of the vapor precursor inside and outside the powder. In fact, the precursor, originating within the powder, has to diffuse towards the substrate and its diffusion coefficient inside the powder is obviously smaller than in the outside Ar:H$_2$ atmosphere. This explains why, right outside of the powder, at its surface, the transport of the volatile specie is enhanced, and hence the thickness of the films is larger than in the buried region.

At larger distances from the powder surface the thickness rapidly decreases because the deposition on the stripe surface gradually consumes the precursor. The thickness gradient can also be followed by the colour of the interference fringes. It can be observed that an appreciable growth occurs until around 3 mm from the powder surface. This experiment suggests that, most probably, the same growth mechanism is responsible for the growth inside and outside the powder, i.e. the final product of the reduction reaction of MoO$_3$ can appear in the same place where the reagents are, or slightly separated from them, giving place to what we are calling here Chemically Driven- CSVT growth (CD-CSVT). One advantage of this growth procedure is that it allows incorporating other sequential process as described below.

c. MoO$_2$ flakes and in-situ tellurization

Flakes are an interesting object in bi-dimensional materials because their large area/volume ratio is an important property for several applications as catalysis or gas sensing. Also, flakes can be transferred to form different electronic devices. As previously discussed, in the described CD-CSVT method, MoO$_2$ growth rate can be controlled by modifying the $d_{s-s}$. It was also observed that at longer distances, the substrate surface is not completely covered by the MoO$_2$ film. This observation suggested the possibility of growing MoO$_2$ nanostructures using this technique.

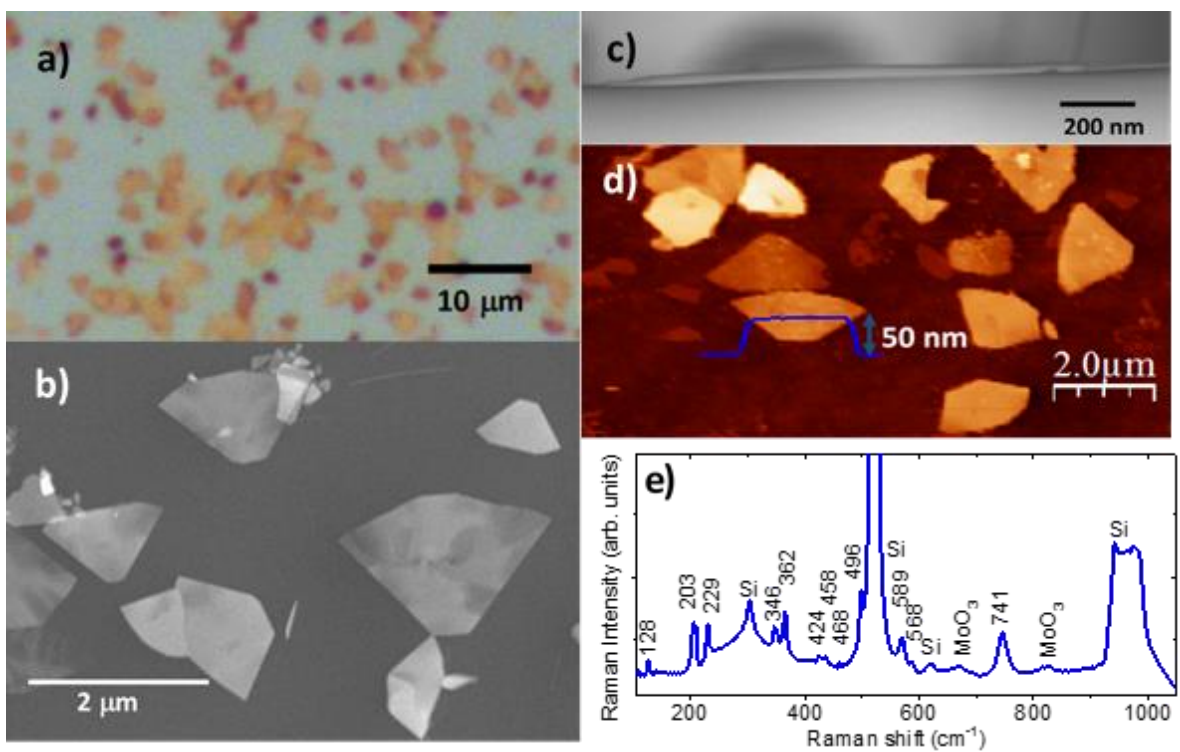

Fig. 6. Optical (a) and SEM (b) top view images of MoO$_2$ flakes. c) SEM cross section of a flake in a cleaved sample. d) AFM image and a height profile of a typical flake; e) Raman spectrum confirming the presence of MoO$_2$.

Fig. 6 shows the formation of flakes for a sample grown at 570 °C for 1h. and a $d_{s\text{-}s}$ of 5 mm. The low growth rate due to the low temperature and the relatively large $d_{s\text{-}s}$ gave rise to the growth of isolated structures in the form of flakes of moderately large dimensions over a Si substrate. An optical microscope view, top and cross section SEM images of such flakes are shown in Fig. 6 a-c). An AFM image including a typical height profile for a flake is shown in Fig. 6.d). The corresponding Raman spectrum in Fig. 6.e) confirms the flakes being MoO$_2$. In this case, in contrast with conformal thick films, a small contribution from the MoO$_3$ phase is observed (features at around 650 and 820 cm$^{-1}$) in the Raman spectrum. This is a consequence of the large surface area in these structures, and the fact that, as verified in XPS measurements, the MoO$_2$ surfaces tend to form MoO$_3$ at environmental conditions. This is an indication that, as in other laminar materials which are not stable in air, MoO$_2$ flakes have to be covered with any protective coating if composition is needed to be preserved over time.

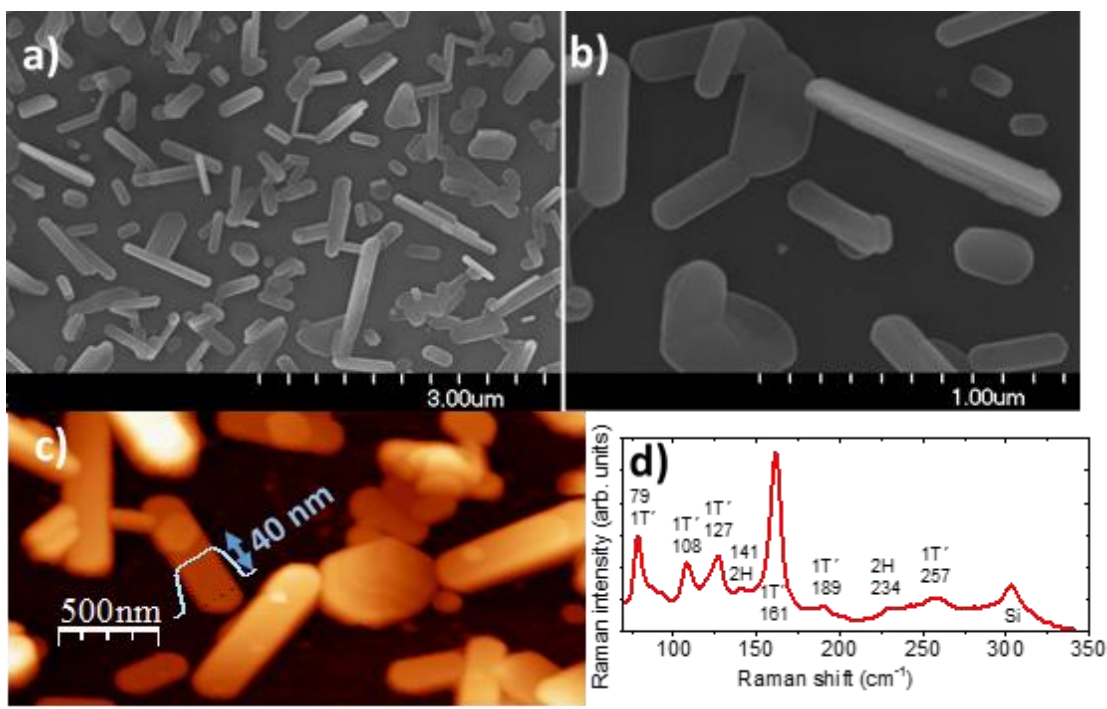

Fig. 7. Top view of MoTe$_2$ obtained by tellurization of an in-situ grown MoO$_2$ at two different magnifications (a, b). c) AFM image and height profile of a typical MoTe$_2$ platelet. d) Raman spectrum for the same sample. 1T´phase with a small contribution of the 2H phase was obtained.

Molybdenum oxides are commonly used as precursors for the growth of molybdenum dichalcogenides. To this end, the molybdenum oxide is typically grown by the sputtering technique and, in a second step, this sputtered oxide is converted to the corresponding dichalcogenide compound by its exposure to the chalcogen element vapors. An interesting application of the present growth method is the possibility of carrying out both the MoO$_2$ growth and its tellurization in a single process. For this task, a graphite boat with an additional bin for the Te source (shown in the supplementary material, see Fig. S.III) was used. Figs. 7a,b) show SEM images of MoTe$_2$ crystals formed in a sequential process comprising an initial step of CD-CSVT growth of the oxide under conditions known to produce MoO$_2$ flakes and a Te vapor annealing in the same boat and at the same temperature. An AFM image with the height profile of a typical crystal is displayed in Fig. 7c). A Raman spectrum in Fig. 7.d) confirms the presence of 1T´MoTe$_2$ with a small component of 2H phase. Since the flakes are relatively thick, if 2D applications are needed, effort has to be devoted to optimize the growth process for obtaining large area and few monolayers thick flakes.

3. Conclusions

A novel chemically driven close space vapor transport (CD-CSVT) technique has been devised for the growth of $MoO_2$ in which, in contrast from traditional CSVT technique, a source/substrate temperature difference is not required for the growth of the films. Instead, the complex reduction reaction of the $MoO_3$ powder, used as source, evolves in such a way that the final $MoO_2$ product appears at the growing surface, located above, a few millimetres away. This mechanism takes place thanks to the presence of an intermediate reaction involving a volatile transport specie, presumably a hydrated or hydroxylated molybdenum oxide. In this sense, different to other growth processes, this mechanism entails only one (complex) reaction that evolves not only in time but also in space. We expect that this CD-CSVT mechanism could be used for thin film growth of other materials presenting complex reactions with volatile intermediate species.

With the CD-CSVT method described here, and controlling growth conditions as the source to substrate distance, a range of films, from relatively compact $MoO_2$ structures to isolated flakes were grown. The possibility to sequentially grow $MoO_2$ and its in-situ tellurization to form $MoTe_2$, was also demonstrated as a first example within the important family of the transition metal dichalcogenides. This represents an efficient alternative to the commonly used tellurization method involving two separated steps: sputtering of Mo oxide and vapour transport of tellurium. We demonstrate the intrinsic instability of $MoO_2$ after atmospheric exposure. Then, an additional advantage of the proposed process is a better control of the precursor Mo oxide composition, since $MoO_2$ is not exposed to air before tellurization. The present procedure can probably be extended to the preparation of other transition metal dichalcogenides as $WTe_2$, $MoSe_2$ or $MoS_2$ for example, and then to the growth of heterostructures combining TMOs or TMDs.

Acknowledgements. OdM, YG and OC thanks the support of the program "Cátedras de Excelencia de la Comunidad de Madrid." This work was partially supported by CAPES-PVE 88881.068066/2014-01 project.